\documentclass{kapproc} 

\InputIfFileExists{psfig.sty}{}{}
\InputIfFileExists{epsf.sty}{}{}

\newcommand{\dd}{ {\textrm d} }

\usepackage{procps} 

\usepackage[dvips]{graphicx}

\upperandlowercase

\nochapequationnumber 

\normallatexbib 

\begin{document}

\articletitle{The inner structure of hybrid stars}

\author{Bela Lukacs, \ Gergely G. Barnafoldi, \ and \  Peter Levai}
\affil{RMKI Research Institute for Particle and Nuclear Physics\\
POB. 49, Budapest, Hungary, H-1525}
\email{lukacs@rmki.kfki.hu, bgergely@rmki.kfki.hu, plevai@rmki.kfki.hu}

\begin{abstract}
Hybrid stars with extremely high central energy density
in their core
are natural laboratories to investigate the appearance and the
properties of compactified extra dimensions with small
compactification radius -- if these extra dimensions exist at all. 
We introduce the necessary formalism to describe quantitatively
these objects and the properties of the formed hydrostatic equilibrium.
Different scenarios of the extra dimensions are discussed
and the characteristic features of these hybrid stars are calculated.
\end{abstract}


\section*{Introduction}
Neutron and quark stars are natural laboratories to investigate the 
interplay of
strong, electro-weak and gravitational interaction. Many
theoretically determined properties of these astrophysical objects
were tested by the observed properties of pulsars, and
detailed calculations exist for these stars\cite{HTWW65,Glend97,Web99,Blas01}.
 
However, if new perspectives appear in the description and understanding
of the gravitational interaction or in the unification of the above
interactions, then revisiting of the models becomes necessary.
Such a reinvestigation was triggered by  the refreshed attention on
compactified extra dimensions\cite{RandSud00}.
Extra dimensions inside neutron stars were investigated
earlier\cite{Liddle90}, but the Kaluza-Klein (K-K) excitation modes
were not considered in the equation of state (EoS). These modes are
important constituents of the recent gravitation theories.
Introducing the K-K modes  into the EoS of fermion
stars at their central core, new features  and properties
emerged~\cite{KanShir}.
 
Here we display a few of our ideas about these extra dimensions,
their possible connection to particle physics and their appearance
in the core of hybrid stars.
We summarize our numerical results
and discuss the observability of extra dimensions in these objects.

\section{The Fifth Dimension}

The introduction of the 5\textsuperscript{th} dimension into the real World
has a long history. 
We do not have any direct information about the extra dimensions, so we have 
an alternative. Either $x^5$ does not exist, or, it is microscopically
small and compact. Obviously in the present paper we take the second
horn of the alternative, for details see Refs.~\cite{lukacs2000,faro2002}.

Quantization puts a serious constraint on five-dimensional motion.
If there is an independence on $x^5$,
then the particle is freely moving in $x^5$.
However, being that direction compactified leads to
an uncertainty in the position with the size of $2 \pi R_c$, where $R_c$
is the compactification radius. Thus a Bohr-type quantum condition appears
formally:
\begin{equation}
        {p^5} = \frac{ n \hbar}{R_c} \ \ .
\end{equation}
Because of the extra motion into the fifth
dimension, an extra mass term appears in 4D descriptions. Considering
compactified radius $R_c \sim 10^{-12}- 10^{-13}$ cm this extra
"mass" is ${\widehat m} \sim 100$ MeV;  similar values appear
in Ref.~\cite{RandSud00}. A quite recent approach of 
hadron spectra by Arkhipov~\cite{Arkhip} is also worthwhile to consult with.

An interesting consequence of the existence of 5\textsuperscript{th} dimension
is that in the "4 dimensional" observations an apparent violation of
the equivalence principle must appear as one can show it by writing the
geodesic equation in 5 dimensions and then projecting it into 4 dimensions.
Without going into details, the $\pm$ sign of $p^5$ causes the appearance
of a "pseudo-charge" ${\widehat q}$ in the 4 dimensional formalism,
\begin{equation}
        {\widehat q} = n \cdot \frac{ 2 \hbar \sqrt{G}}{c\,\,R_c} \ \ ,
\end{equation}
which acts in a vector-scalar interaction. We can directly see that 
${\widehat q}$ is {\it not} the electric charge. Indeed 
${\widehat q}^2 < 16 \pi G m_0^2 $, where
$G$ is the gravitational constant and $m_0$ is the rest mass~\cite{LukPach85}.
This is either some familiar quantum number (e.g. strangeness, $S$), or 
some other not yet observed charge.

\section{Field Equations}

We are interested in final states of stellar evolution. 
Therefore we can restrict ourselves to static configurations.
Also, fluid-like behavior seems appropriate {\it in the microscopical
dimensions}. 
Therefore we are looking for {\it static} configurations. Also, some
fluid-like behaviour is expected in the sense that stresses in the {\it
macroscopical} directions freely equilibrate. Then in 3 spatial directions
{\it isotropy} is expected and thence {\it spherical symmetry}. Finally, 
in the lack of any information so far, we may assume symmetry in the extra
dimension. Then in proper GR language we are looking for solutions with 5
Killing vectors
$$K_{Ai;k}+K_{Ak;i}=0$$
$A=0,1,2,3,5,$ with the commutation relations
\begin{equation}
[K_0,K_B]=0, \ \ \ \ [K_5,K_B]=0, \ \ \ \ 
[K_{\alpha},K_{\beta}]={\epsilon}_{\alpha \beta \gamma }K_{\gamma}
\end{equation}
so that the subgroup ${1,2,3}$ acts with 2 dimensional transitivity. Hence the
metric has the unique form
\begin{eqnarray}
\dd s^2&=&
e^{2\nu} \, \, \dd t^2 - e^{2\lambda} \, \, 
\dd r^2 - r^2 \dd {\Omega}^2-e^{2\Phi }({\dd x^5})^2 \ \ ,
\end{eqnarray}
where the quantities $\nu$ , $\lambda$ and $\Phi$ depend only on $r$.
As for the energy-momentum tensor we get
\begin{eqnarray}
T^{ik} & =&
{\rm diag} \  ( \varepsilon \ e^{2 \nu},  \ P \ e^{2 \lambda},
 \ P\ r^2, \ P\  r^2 \sin^2 \theta, \ P_5 \ e^{2 \Phi} ) \ .
\end{eqnarray}

Total spatial equipartition of particle momenta is not expected
(so $P \neq P_5$); indeed we shall see that it does not happen generally.

The Einstein equations are read as
\begin{eqnarray}
{ -\gamma \ \varepsilon }
&{ =}&
{ e^{-2 \lambda} \left[ { \Phi'' + \Phi'^2 - \lambda' \Phi'
+ \frac{2 \Phi'}{r}} - \frac{2 \lambda'}{r} + \frac{1}{r^2} \right]
- \frac{1}{r^2} }  \label{einst1} \\
{ -\gamma \ P }
&{=}&
{ e^{-2 \lambda} \left[ { - \nu' \Phi'
- \frac{2 \Phi'}{r}} - \frac{2 \nu'}{r} - \frac{1}{r^2} \right]
+ \frac{1}{r^2} } \label{einst2}  \\
{ -\gamma \ P }
&{ =}&
{ e^{-2 \lambda} \left[ -\nu'' -\nu'^2 + \nu' \lambda'
{ - \Phi'' - \Phi'^2 -\nu' \Phi' + \lambda' \Phi'  - } \right.} \nonumber \\
&& \hspace*{5.5truecm} { \left. {
- \frac{2 \Phi'}{r}}  - \frac{\nu'}{r} + \frac{2 \lambda'}{r} \right]
 }  \label{einst3} \\
{ -\gamma \ P_5 }
&{ =} &
{ e^{-2 \lambda} \left[  -\nu'' -\nu'^2 + \nu' \lambda'
 - \frac{2\nu'}{r}  + \frac{2\lambda'}{r} - \frac{1}{r^2} \right]
+ \frac{1}{r^2} }   \label{einst4}
\end{eqnarray}
where $\gamma = 8 \pi G / c^4$ and all quantities can depend only on the radius $r$.

For final states ($T=0$) all local material 
characteristics of the fluid are expected
to depend on one thermodynamic quantity, say on particle density $n$.
The material equations,
\begin{equation}
\varepsilon = \varepsilon(n); \  P=P(n); \  P_5 = P_5(n) \ ,
\end{equation}
and the Einstein equations in eqs.~(\ref{einst1})-(\ref{einst4}) 
are all independent equations, all further equations are consequences.
(Indeed there are some thermodynamical constraints between 
$\varepsilon, P$ and $P_5$.) Considering the appropriate
Bianchi identity, one obtains
\begin{equation}
T^{ir}_{\,\,\,\,\,;r} = 0 \ \ \longrightarrow \ \ 
P' = -\nu '(\varepsilon + P) + (P_5-P) \Phi' \ .
\end{equation}
This equation clearly demonstrates the influence of the extra dimensional 
behavior on the normal pressure, $P=P_1=P_2=P_3$.

The Einstein equations eqs.~(\ref{einst1})-(\ref{einst4}) 
contain two extra variables compared to the more familiar
4 dimensional case, namely $P_5$ and $\Phi$. However, $P_5(n)$ is a known
function of the particle number density and specified by the
actual interaction in the matter. Thus $\Phi(r)$ is the only new degree 
of freedom determined by the extra equation.

\section{A Special Solution of the Field Equations}

For specially chosen pressure $P_5$ there is a unique solution of 
the Einstein equations in eqs.~(\ref{einst1})-(\ref{einst4}), namely $\Phi'=0$. 
In this case eqs.~(\ref{einst1})-(\ref{einst3}) 
can be solved separately with $\Phi=0$ and then
the last eq.~(\ref{einst4}) gives $P_5$. 
Although these solutions do not differ formally from the 
4 dimensional neutron star solution (except for $P_5$),
but the extra dimension will have its influence
on $\varepsilon(n)$ and $P(n)$~\cite{KanShir}.

Let us start with a single massive fermion ("neutron", $N$).
Since the minimal nonzero fifth momentum component is 
$\vert p^5 \vert = \hbar/R_c$, then the extra direction of the phase space
is not populated until the Fermi-momentum $p_F < \hbar/R_c$. 
However, at the threshold both
$p^5 = \pm \hbar /R_c$ states appear. They mimic another
("excited") particle with mass $m_X = \sqrt{m_N^2 + (\hbar/R_c )^2 }$
(with a nonelectric "charge" 
${\widehat q} = \pm \frac{2 \hbar \sqrt{G}}{c R_c}$ as well).
The equations obtain a form as if this second particle also appeared in
complete thermodynamical equilibrium with the neutron: $\mu_X = \mu_N$.
This phenomena is repeated in any case, when $p$ exceeds 
a threshold $n\hbar / R_c$.

\vspace*{-5mm}
\begin{figure}[htb]
\begin{minipage}[t]{6.5cm} {\epsfxsize=6cm \epsfbox{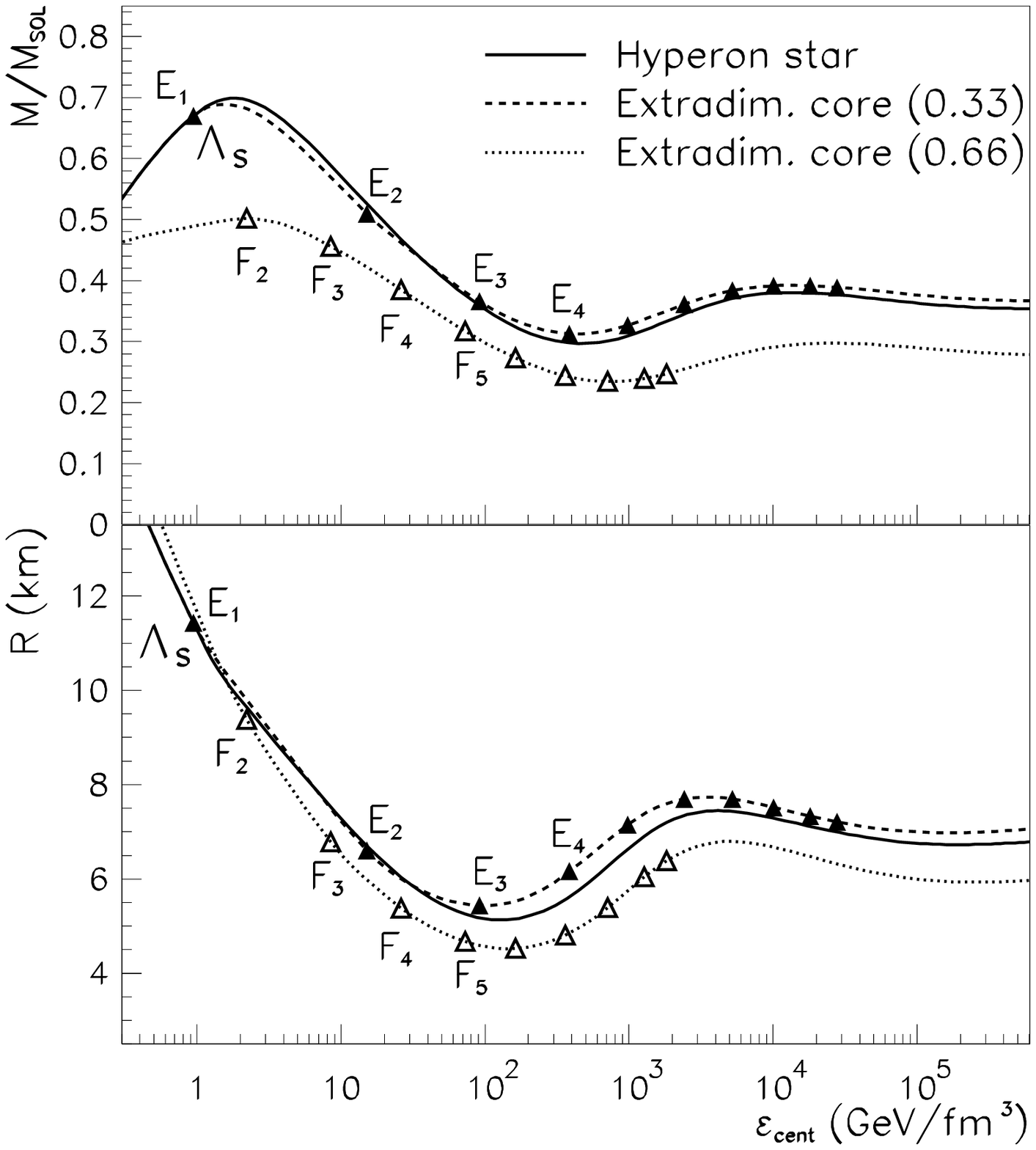}}
\end{minipage} \hfill
\vspace*{- 67mm} \hspace*{6.0cm} \hfill
\begin{minipage}[t]{6.5cm}
{\epsfxsize=6.0cm \epsfbox{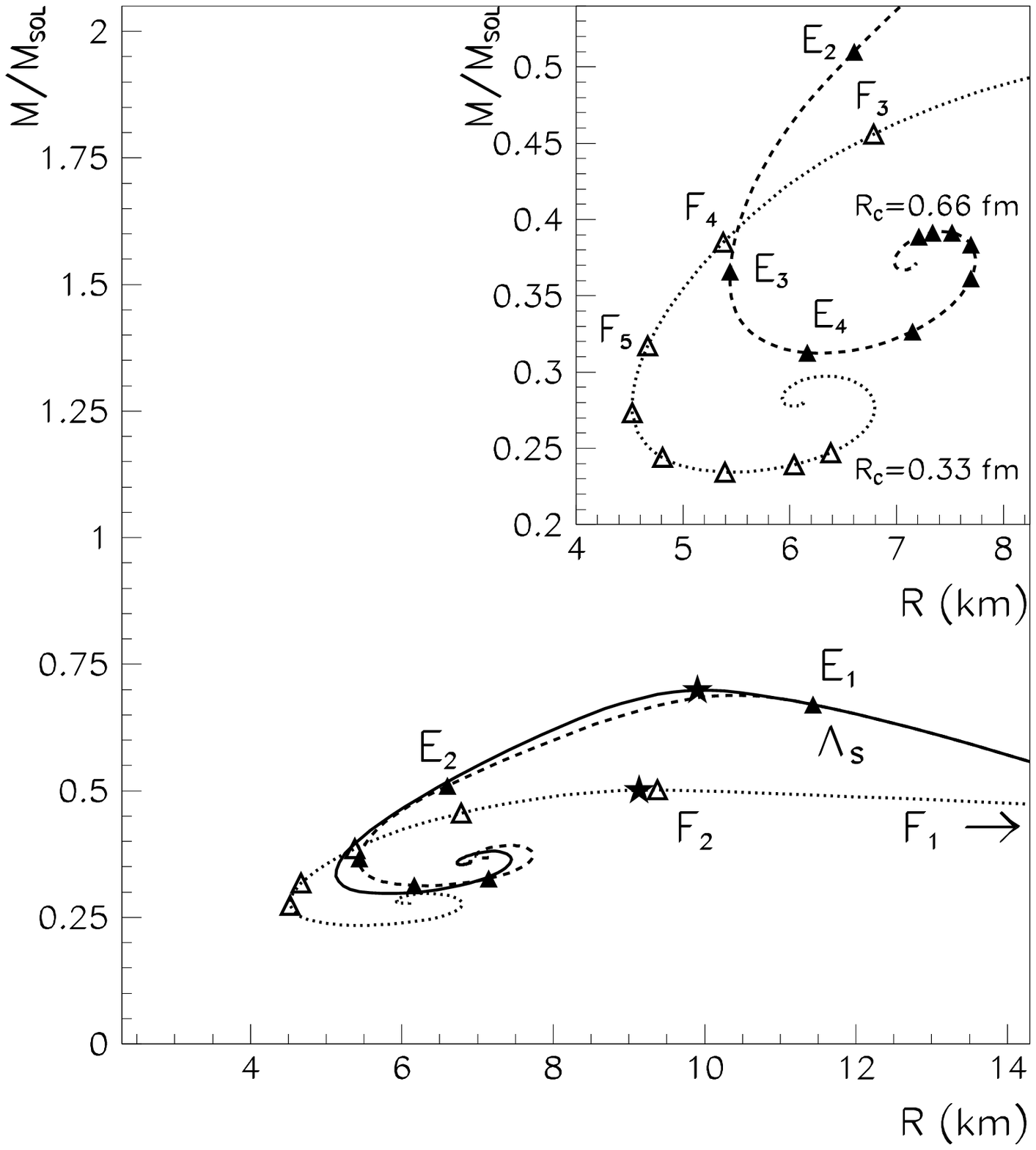}}
\end{minipage}
\vspace*{-8mm}
\caption{The mass and the radius of heavy hadron stars in 4 dimensions
and neutron stars including extradimensional K-K modes into the
core (from Ref.~\cite{faro2002}).
\label{fig1}}
\end{figure}

Fig. 1. compares two calculations. The solid line is 4 dimensional calculation
with neutrons and $\Lambda$ hyperons, the lightest neutral strange baryon. 
(Ambartsumyan and Saakyan calculated such hybrid compact stars
as far back as in 1960~\cite{Ambart}.) 
The dashed and dotted lines come from 
5-dimensional calculations with only neutrons, but moving in the
extra dimension, too. The higher excitations are started from the
triangles and they are named as $E_1$, $E_2$, ..., at compactified radius
$R_c=0.33$ fm. In this case the 5 dimensional neutron 
star is almost indistinguishable from a neutron
star with $\Lambda$ core.
However, choosing another compactification radius (e.g. $R_c = 0.66$ fm)
one obtains reasonable differences, which can be seen on Fig. 1.
The $F_2$, $F_3$, ..., indicate the appearance of the excited modes in the
latter case ($F_1$ was left outside of the figure).

\section{On the General Solution of the Field Equations}

In the most generic case it is true that the derivative
$\Phi' \neq 0$. The interior solution of eqs.~(\ref{einst1})-(\ref{einst4})
can be obtained only numerically. In principle there is no problem
to evaluate these equations, but the structure of them  
must be understood.

First, there are constraints among the material quantities,
which are rather straightforward for ideal quantum fluids.
In case of interacting matter one can recall the 
self-consistent description of the interacting fermions in
the 4 dimensional analogy~\cite{Zim88}. 
Second, the particle number density in the center, $n(r=0)$, 
is a free initial condition, as it was in the 4 dimensional case.
Instead of density, we may use energy density,
$\varepsilon(r=0)\equiv  \varepsilon_0$, as well.

Third, the metric tensor is determined by the variables $\Phi, \nu, \lambda$.
On the other hand $\Phi$ and $\nu$
never appear  in eqs.(\ref{einst1})-(\ref{einst4})
(reflecting the fact that $x^0$ and $x^5$ constant
dilatations are always possible without harming the commutator relations
for the Killing motions), so these equations are of
first order on $\Phi'$,  $\nu'$ and $\lambda$.
However, the equations can be rearranged resulting in the following 
symbolic structure:
\begin{eqnarray}
\Phi' &=& \Phi'(\lambda', \lambda, n) \label{form1} \\
\nu' &=& \nu' (\lambda', \lambda, n) \label{form2} \\
\lambda'' &=& \lambda'' (\lambda', \lambda, n) \label{form3}
\end{eqnarray}
The eqs.~(\ref{form1})-(\ref{form2}) are algebraic equations for
$\Phi'$ and $\nu'$, but
eq.~(\ref{form3}) is a differential equation of second order and
it should be reorganized formally into 2 equations of first order.

Fourth,  initial conditions are needed for $\Phi$, $\nu$,
$\lambda$ and $\lambda'$.
For the first two they are free additive constants and for $\lambda$
and $\lambda'$ we can proceed as in 4 dimensions.
Anyway, it is clear that even if we choose such initial conditions that
\mbox{$\Phi'(r=0)=0$}, then it does not remain zero out of the center.

\section{Matching Conditions on the Surface}

The integration must go until the fluid-vacuum interface at $r_s$,
where an exterior vacuum solution continues the interior one. 
The details of matching conditions can be found in Ref.~\cite{lich55}.
Applying them on the present problem we get the continuity of certain
derivatives of the metric tensor, and those for the energy-momentum tensor
result in the single equation,
\begin{equation}
P(r_s) = 0 \ .
\end{equation}
Observe that no constraint is obtained either for
$\varepsilon$ or $P_5$.

Although $P_5$ might be anything at $r_s$ from the matching conditions,
it will be zero for all general (not very exotic) systems.
Namely, $P=0$ is expected at some {\it low} particle number density
$n_s$ on the surface, which is generally much below $p_5 = \hbar / R_c$.
So motions in the $5^{th}$ dimension cease already somewhere in the
interior of the star.

Henceforth vacuum solutions of eqs.~(\ref{einst1})-(\ref{einst4})
are valid until infinity, and $\Phi(r_s)$, $\nu(r_s)$,
$\lambda(r_s)$, and  $\lambda'(r_s)$
 give initial conditions for the external solution.
Interestingly enough, the external solutions can be obtained in analytical
(albeit partly implicit) forms.

\section{The External Solution}

The external solution of the Einstein equations can be obtained in
analytic form. Equaling the matter contribution with zero 
on the left hand sides and using the new variables $\alpha$ and $\beta$ as 
\begin{eqnarray}
\nu '  &\equiv & \alpha + \beta \label{albe1} \\
\Phi ' &\equiv & \alpha - \beta \ ,  \label{albe2}
\end{eqnarray}
the 4 Einstein equations  results in 3 differential equations,
\begin{eqnarray}
\lambda ' &=& {1 \over {r}} (1 - e^{2 \lambda}) + 2 \alpha  \label{lambda} \\
\psi ' &=& - {{\psi}\over {r^2}} (1 + e^{2 \lambda}) \label{psi} \ ,
\end{eqnarray}
(where $\psi \equiv \alpha, \beta$), and one more equation
becomes an  identity.
Eqs.~(\ref{lambda})-(\ref{psi}) can be integrated as
\begin{eqnarray}
\alpha &=& {{A} \over {r}} e^{-Y} \\
\beta  &=& {{B} \over {r}} e^{-Y} 
\end{eqnarray}
with two constants $A,B$. For variable $Y$ eq.(\ref{lambda}) yields 
the following:
\begin{equation}
Y ' = {1 \over r} \left[ (A^2 + B^2) e^{-2 Y} + 4 A e^{-Y} +1 \right]
\end{equation}
This last equation can be solved in an implicit form. There are
three solutions according to the sign of the determinant.
Here we give the one for positive determinant:
\begin{equation}
\ln \left[ {{r}\over {r_0}} \right] 
= \ln \left[ \left({{y-y_+}\over {y_0 - y_+}} \right)^u
                   \left({{y-y_-}\over {y_0 - y_-}} \right)^v
            \right]
\end{equation}
where $r_0$ and $y_0$ are constants of integration. Furthermore,
\begin{eqnarray}
D &=& \sqrt{3 A^2 - B^2} \\
y_{\pm} &=& -2A \pm {D} \\
u &=& { {-2A + {D}}\over {2 {D}} } \\
v &=& { {+2A + {D}}\over {2 {D}} } 
\end{eqnarray}
and then
\begin{eqnarray}
Y &=& \ln y \\
e^{2 \lambda} &=& (A^2 + B^2) e^{-2 Y} + 4 A e^{-Y} + 1
\end{eqnarray}
Of course, this solution goes to flat space-time as 
$r\longrightarrow \infty$.

The $\Phi ' = 0$ special solution belongs to $A=B$.
However, as we claimed earlier, $\Phi '(r_s) \neq 0$.

\section{Conclusion}

We do not know compact stars close to Sun, so the observations are rather hazy
for details. However some data exist for masses and radii. In the previous
Chapters we showed that the calculations are as straightforward in 5
dimensions as in 4. While some problems must be relegated to the next, last
Chapter, in principle one could compare observations with 4- and 5-dimensional
calculations, and try to decide if space is 3 dimensional or 4. The difference
comes from two reasons.
 
1) At some Fermi momentum the phase space opens up in the 
5\textsuperscript{th} direction.
Henceforth $p_F$ increases slower, which will be reflected by the M(R)
relation. However note that 5\textsuperscript{th} dimensional 
effects can mimic particle excitations.
 
2) Quite new effects can, in principle, be observed around compact stars. Of
course, they may be moderate and the present observational techniques are
probably insufficient. However, consider 2 situations:

You can perform light deflection experiments. 
Trajectories differ in 4 and 5 dimensions.
 
You may perform "free fall" experiments in vacuum but not very far from the
star. Then 1) particles moving also into the extra direction would fall
"anomalously", but this is a simple Eotvos-type experiment and can be done in
lab too. However 2) $\Phi ' \neq 0$, so also a "scalar force" appears for
particles moving in the extra dimension, although we are outside of the exotic
interior of the star. This means that accelerator experiments not too far from
compact stars would show up surprising results 
{\it if} 5\textsuperscript{th} Dimension exists. 
So not only the dense source would be different. Gravity is a phenomenon of
Space-time and 5-dimensional Space-Time differs from the 4-dimensional one.
 
Before truly realistic 5-dimensional calculations some problems must be fully
clarified. They are listed in the last Chapter.

\section{Outlook}

While we can calculate 5-dimensional compact stars in this moment,
the results still depend on parameters to be fixed in the future,
We, somewhat arbitrarily, classify problems into 3 groups:
factual, technical and fundamental.

1) Factual problems: \\ \hspace*{0.4truecm}
{\it   The value of the compactification radius, $R_c$:} \ 
In the present approach this radius was a free parameter.
For demonstration we chose the radius $R_c=0.33 \cdot 10^{-13} $ cm,
when the strange $\Lambda$ baryon could behave as the first excitation 
of a neutron. Such an extradimensional object can mimics a compact
star with neutrons in the mantle and $\Lambda$'s in the core. With smaller
$R_c$ the "exotic" component appears at larger densities -- we may run
into the unstable region of the hybrid star and the extra dimension
remains undetectable. However, with larger $R_c$ the mass gap 
becomes smaller and 
the ``transition'' happens at "familiar" neutron star
densities. In this way, reliable observations could lead to an upper 
bound on $R_c$. 

{\it The quantum number connected with motion in $x^5$:} \
We saw that a neutron, starting to move in the 
5\textsuperscript{th} direction, appears in macroscopic
4-dimensional description as if a neutral baryon,  but
with a higher mass. Particle tables mention such particles in abundance. But
we also mentioned that 5 dimensional GR gives that the 4 dimensional
projection of the geodesics of such particles would exhibit a really weak
apparent violation of the equivalence principle, via a vector-scalar force. In
first approximation this would mimic a Coulomb-{\it like} force, but cca.
$10^{-39}$ times weaker. Now, there was one serious attempt to see the
violation of the equivalence principle 
(in the Eotvos experiment~\cite{Fish86} );
there it seemed that the strength was in this range. The tentative idea was
that the violation might be connected with hypercharge 
$Y=B+S$, and there was also
a discussion~\cite{Fish91} 
whether the mass difference of $K^0$ and anti-$K^0$ came from
the interaction of Earth's particles with the kaon in lab. Now, in some sense
the "first excitation" of $n$ in $Y$ is $\Lambda$. It would be a nice minimal
theory if Fifth Dimension would explain the CP-violation of weak interaction
as well, but of course we cannot expect this.

{\it The way of compactification:} \ 
Here we chose the simplest
compactification: $x^5 \equiv x^5 +2\pi R_c$. This is a cylindrical
compactification. However more complicated cases are also possible, albeit not
arbitrary ones. A branch of GR  lists all the cases; now we mention only the
2-dimensional example that from a plane by cut and sew you can produce of
course a compact cylinder, but the mantle of a cone as well.

{\it The number of the extra dimensions:}  \
 The present example was the
simplest nontrivial case. However maybe particle physicists would prefer 6
extra dimensions appearing with the same length scale in some group structure.
The "extra forces" appearing from Fifth Dimension are
gravitational, the extra quantum number appearing here cannot be the source
of QCD forces~\cite{LukPach85}. 
In Ref.~\cite{faro2002} we discussed  {\it two} extra dimensions with
different scales.

2) Technical problems: \\ \hspace*{0.4truecm}
{\it Initial conditions:} \ We mentioned that we need to fix 3 initial
conditions in the center: one for the central density of matter ($n(0)=n_0$ or
$\epsilon(0)=\epsilon_0$), and two for the metric, either for $\lambda(0)$ and
$\lambda '(0)$, or, equivalently, for $\lambda(0)$ and
$\Phi '(0)$. But we cannot properly impose these conditions in $r=0$, and 
these conditions are somehow not independent. However the technical problem is
well known already in 4 dimensions~\cite{HTWW65} . 
First, the proper way is to
approximate the innermost core of radius $\delta $ with a homogeneous sphere
of density $n_0$, where the exact value of $\delta $ is irrelevant if small
enough. Then $n=n_0$ at $r=\delta$, and 
$e^{2\lambda}=1-8\pi {\delta }^3\epsilon_0/3c^2$ there.

{\it Interactions:} \ $N-N$ interaction can be taken
into account the same way as in Ref.\cite{Zim88}. 
Neutrons moving into the extra dimension  may interact differently
because of an extra momentum dependence. Such momentum dependence appears
e.g. in Walecka-type~\cite{Wal74,Zim90} 
construction.

3) Theory: \\ \hspace*{0.4truecm}
A consequent 5-dimensional treatment would require Unified Theory of Quantum
Mechanics and General Relativity. This unified theory is not available now,
and we know evidences that present QM is incompatible with present GR. 
The well-known demonstrative examples are generally between QFT and GR (e.g.
the notion of Quantum Field Theory vacua is only Lorentz-invariant and hence
come ambiguities about the existence of cosmological 
Hawking radiations~\cite{GW}).
But also, it is a fundamental problem that the lhs. of Einstein equation is
$c$-number, while the rhs. should be a quantum object.

In the present case QM and GR have to be used in a compatible way for the
proper "Bohr-type" quantization of $p^5$. Our chosen way was intuitive. Sure,
it is correct for $\Phi '=0$. Then one can always introduce such a coordinate
system where $\Phi=0$ and then the circumference is the usual $2\pi R_c$. But
in the general case one might use $2\pi R_c$ in the condition as well as
$2\pi e^\Phi R_c$. All suggestions up to now (as. e.g. Ref.~\cite{Souriau}) 
are restricted to the simpler case.

\newpage

This paper has displayed the introduction of 
5\textsuperscript{th} dimension into astrophysical description in the framework
of GR. The new degrees of freedom may solve old problems, but
open new questions, as well. We think that this attempt is worthwhile.

\begin{acknowledgments}
One of the author (B.L.) would like to thank for the organizers of this
conference for their warm hospitality.
This work was supported by the MTA-JINR Grant and OTKA T043455.
\end{acknowledgments}


\begin{chapthebibliography}{99}
\bibitem{HTWW65} B.K. Harrison, K.S. Thorne, M. Wakano, and J.A. Wheeler.
{\it ``Gravitation theory and gravitational collapse''},
{\rm University of Chicago Press}, 1965.
 
\bibitem{Glend97} N.K. Glendenning. {\it ``Compact Stars''}, 
{\rm Springer}, 1997;
and references therein.
 
\bibitem{Web99} F. Weber. {\it ``Astrophysical Laboratories for Nuclear
   and Particle Physics''}, {\rm  IOP Publishing}, Bristol, 1999.
 
\bibitem{Blas01} D. Blaschke, N.K. Glendenning, and A. Sedrakian (Eds.),
{\it ``Physics of neutron star interiors''}, {\it Springer}, Heidelberg, 2001.
 
\bibitem{RandSud00}L. Randall and R. Sudrum.
  {\it Phys. Rev. Lett.} {\bf 83}, 3370 (1999); {\it ibid.} 4690.
 
\bibitem{Liddle90} A.R. Liddle {\rm et al.}.
                   {\it Class. Quantum Grav.} {\bf 7}, 1009 (1990).
 
\bibitem{KanShir} N. Kan and K. Shiraishi,
     {\it Phys. Rev.} {\bf D66}, 105014 (2002).

\bibitem{lukacs2000}
      B. Lukacs, {\it ``May Kaluza ride again?''},
      in Relativity Today, eds. C. Hoenselaers, Z. Perjes,
      Akademiai Kiado, Budapest, 2000, p.161.

\bibitem{faro2002}
      G.G. Barnafoldi, P. Levai, and B. Lukacs,
      {\it ``Heavy quarks or compactified extra dimensions
       in the core of hybrid stars''},
      in Proceedings of the 4th Int. Workshop on New Worlds in
      Astroparticle Physics, Faro, Portugal,
      Worlds Scientific, Singapoure, 2003.
      ({\tt astro-ph/0312330}).

\bibitem{LukPach85} B. Lukacs and T. Pacher,
 {\it Phys. Lett.} {\bf A113}, 200 (1985).


\bibitem{Arkhip} A.A. Arkhipov, 
{\it ``Hadronic Spectra and Kaluza-Klein picture of the World''}
(hep-ph/0309327) and references therein.

\bibitem{Ambart} V.A. Ambartsumyan, G.S. Saakyan,
   {\it Astron. Zh.} {\bf 37}, 193 (1960).

\bibitem{Zim88}
 J. Zimanyi, B. Lukacs, P. Levai, et al.
{\it Nucl. Phys.} {\bf A484}, 647 (1988).

\bibitem{lich55}
  A. Lichnerowicz, {\it ``Theories relativistes de la gravitation at
  d'electromagnetisme''}, Masson, Paris, 1955. 

\bibitem{Fish86}
  E. Fischbach et al., {\it Phys. Rev. Lett.} {\bf 56}, 3 (1986).

\bibitem{Fish91}
  S.H. Aronson et al., {\it Phy. Rev. Lett.} {\bf 56}, 1342 (1986).

\bibitem{Wal74}
  J.D. Walecka, 
 {\it Ann. Phys.} {\bf 83}, 491 (1974).

\bibitem{Zim90}
  J. Zimanyi and S.A. Moszkowski,
  {\it Phys. Rev.} {\bf C42}, 1416 (1990).
  

\bibitem{GW}
  G.W. Gibbons and S. Hawking, 
  {\it Phys. Rev.} {\bf D15}, 2738 (1977).

\bibitem{Souriau}
  J.-M. Souriau, {\it Nuovo Cim.} {\bf 30}, 565 (1963).

\end{chapthebibliography}
\end{document}